\documentclass[showpacs,twocolumn,aps]{revtex4}
\emergencystretch=\maxdimen
\usepackage{dcolumn}
\usepackage{amsmath,amssymb,mathrsfs}
\usepackage[latin1]{inputenc}
\usepackage{graphicx, psfrag}
\usepackage[colorlinks=true, citecolor=blue, urlcolor = blue, linkcolor= red, bookmarks=true]{hyperref}
\usepackage{float}
\usepackage{amsfonts}
\usepackage{dcolumn}
\usepackage{hyperref}
\usepackage{subfigure}
\usepackage{pgfplots}
\usepackage{epstopdf}
\usepackage{booktabs}
\usepackage{gensymb}

\makeatletter
\def\btt#1{\texttt{\@backslashchar#1}}
\DeclareRobustCommand\bblash{\btt{\@backslashchar}} \makeatother

\begin{document}
	
\title[]{Gravitational lensing by  charged black hole in regularized $4D$  Einstein-Gauss-Bonnet gravity}
\author{Rahul Kumar$^{a}$}\email{rahul.phy3@gmail.com}
\author{Shafqat Ul Islam $^{a}$}\email{Shafphy@gmail.com}
\author{Sushant~G.~Ghosh$^{a,\;b}$} \email{sghosh2@jmi.ac.in}
	
\affiliation{$^{a}$ Centre for Theoretical Physics, Jamia Millia Islamia, New Delhi 110025, India}
\affiliation{$^{b}$ Astrophysics and Cosmology Research Unit, School of Mathematics, Statistics and Computer Science, University of KwaZulu-Natal, Private Bag 54001, Durban 4000, South Africa}
	
\date{\today}

\begin{abstract}
Among the higher curvature gravities, the most extensively studied theory is the so-called Einstein-Gauss-Bonnet (EGB) gravity, whose Lagrangian contains Einstein term with the GB combination of quadratic curvature terms, and the GB term yields nontrivial gravitational dynamics in $ D\geq5$. Recently there has been a surge of interest in regularizing, a $ D \to 4 $ limit of,  the EGB gravity, and the resulting regularized $4D$ EGB gravity valid in $4D$. We consider gravitational lensing by Charged black holes in the  $4D$ EGB gravity theory to calculate the light deflection coefficients in strong-field limits $\bar{a}$ and $\bar{b}$, while former increases with increasing  GB parameter  $\alpha$ and charge $q$, later decrease. We also find a decrease in the deflection angle $\alpha_D$, angular position $\theta_{\infty}$ decreases more slowly and impact parameter for photon orbits $u_{m}$ more quickly, but angular separation $s$ increases more rapidly with $\alpha$ and charge $q$. We compare our results with those for analogous black holes in General Relativity (GR) and also the formalism is applied to discuss the astrophysical consequences in the case of the supermassive black holes Sgr A* and M87*.
\end{abstract}
	
\maketitle

\section{Introduction}
General Relativity (GR) not only predicts the existence of black holes but also a mean to observe them through the gravitational impact on the electromagnetic radiation moving in the near vicinity of black holes. The advent of horizon-scale observations of astrophysical black holes \cite{Doeleman:2008qh,Akiyama:2019cqa} offer an unprecedented opportunity to understand the intricate details of photon geodesics in the black hole spacetimes \cite{Carter:1968rr,Gralla:2019drh}, which has become of physical relevance in present-day astronomy. Photons passing in the gravitational field of a compact astrophysical object get to deviate from their original path and the phenomenon is known as the gravitational lensing \cite{Einstein:1956zz,Synge:1966,Darwin}. Though the theory of gravitational lensing was primarily developed in the weak-field thin-lens approximation for small deflection angles \cite{Liebes:1964zz,Schneider}, but in black hole spacetimes, where photons can travel close to the gravitational radius, a full treatment of lensing theory valid even in strong-field gravity regime is required. Darwin \cite{Darwin} led the study of strong gravitational lensing theory, and later Virbhadra and Ellis \cite{Virbhadra:1999nm} numerically calculated the deflection angle due to the Schwarzschild black hole in an asymptotically flat background. Using an alternative formulation, Frittelli, Kling and Newman \cite{Frittelli:1999yf} analytically obtained an exact lens equation. A significant interest in the strong gravitational lensing developed by the Bozza et al. \cite{Bozza:2001xd}, who gave a general and systematic investigation of light bending in the strong-gravity region, and exploiting the source-lens-observer geometry obtained the analytical expressions for the source's images positions. 

One of the generic features of strong gravitational lensing is the logarithmic divergence of the deflection angle in the impact parameter and the existence of relativistic images produced due to multiple winding of light around the black hole before emanating in observer's direction \cite{Bozza:2002zj}. The strong gravitational lensing relevance for predicting the strong-field features of gravity, testing and comparing various theories of gravity in the strong-field regime, estimating black hole parameters, and deducing nature of any matter distributions in black hole background has resulted in a vast, comprehensive literature \cite{Will,Bhadra:2003zs,Bozza:2007gt,Virbhadra:2007kw,Wei:2011nj}. Also, the gravitational lensing for various modifications of Schwarzschild geometry arising due to modified gravities, e.g., regular black holes  \cite{Eiroa:2010wm,Ovgun:2019wej}, massive gravity black holes \cite{Panpanich:2019mll}, $ f(R) $ black holes  \cite{frlens} and Einstein-Gauss-Bonnet (EGB) gravity models \cite{egblens} have been investigated.

EGB gravity theory is a natural extension of GR to higher dimensions $D\geq 5$, in which Lagrangian density admits quadratic corrections constructed from the curvature tensors invariants \cite{Lanczos:1938sf,Lovelock:1971yv}. The EGB gravity, which naturally appears in the low-energy limit of string theory \cite{Zwiebach:1985uq}, preserves the degrees of freedom and is free from gravitational instabilities and thereby leads to the ghost-free nontrivial gravitational self-interactions \cite{Nojiri:2018ouv}. Due to much broader theoretical setup and consistency with the available astrophysical data, the EGB gravity is subject of intense research in varieties of context over the past decades \cite{Boulware:1985wk,Ghosh1:2018bxg,Koivisto:2006xf}.

The Gauss-Bonnet (GB) correction to the Einstein-Hilbert action is a topological invariant in $D=4$ and therefore does not make any contribution to the gravitational dynamics. This issue of $4D$ regularization of EGB gravity is revived recently by Glavan and Lin \cite{Glavan:2019inb}, who by re-scaling the GB coupling parameter as $\alpha\to \alpha/(D-4)$ defined the $4D$ theory as the limit of $D\to 4$ at the level of field's equation. Tomozawa \cite{ Tomozawa:2011gp} earlier proposed this kind of regularization as quantum corrections to Einstein gravity, and they also found the spherically symmetric black hole solution. Later  Cognola {\it et al.} \cite{Cognola:2013fva}   gave a simplified approach for Tomozawa \cite{Tomozawa:2011gp} formulation, which  mimic quantum corrections due to a GB invariant within a classical Lagrangian formalism. Further, the static and spherically symmetric black hole solution \cite{Tomozawa:2011gp,Cognola:2013fva,Glavan:2019inb} of $4D$ EGB gravity is identical as those found in semi-classical Einstein's equations with conformal anomaly \cite{Cai:2009ua}, regularized Lovelock gravity \cite{Casalino:2020kbt,Konoplya:2020qqh}, and the Horndeski scalar-tensor theory \cite{Lu:2020iav}.  

Hence, the $4D$ EGB gravity witnessed significant attentions that  includes finding black hole solutions and  investigating  their properties \cite{Fernandes:2020rpa, Konoplya:2020juj,Singh:2020nwo}, Vaidya-like solution \cite{Vaidya}, black holes coupled with magnetic charge \cite{regular}, and also rotating black holes \cite{ShaRef}. Other probes include gravitational lensing \cite{lensing}, black hole shadows, \cite{Guo:2020zmf,Konoplya:2020bxa,ShaRef}, derivation of regularized field equations \cite{Fernandes:2020nbq},  Morris-Thorne-like wormholes \cite{Jusufi:2020yus}, and black hole thermodynamics \cite{HosseiniMansoori:2020yfj}. Nonetheless, several questions \cite{Ai:2020peo,Hennigar:2020lsl,Shu:2020cjw,Gurses:2020ofy,Mahapatra:2020rds} have been raised on the procedure adapted in \cite{Glavan:2019inb}, and also some remedies have been suggested to overcome \cite{Lu:2020iav,Kobayashi:2020wqy,Hennigar:2020lsl,Casalino:2020kbt,Ma:2020ufk,Arrechea:2020evj,Aoki:2020lig}. However, it turns out that the spherically symmetric $4D$ black hole solution found in \cite{Cognola:2013fva,Glavan:2019inb} remains valid for these regularized theories \cite{Lu:2020iav,Hennigar:2020lsl,Casalino:2020kbt,Fernandes:2020nbq,Ma:2020ufk}, but may not beyond spherical symmetry \cite{Hennigar:2020lsl}. It was argued in \cite{Glavan:2019inb}, without an explicit proof, that a physical observer could never reach the curvature singularity of $4D$ EGB black hole given the repulsive effect of gravity at short distances. However, later considering the geodesics equations, this claim was refuted by Arrechea \textit{et al. }\cite{Arrechea:2020evj}. The infalling observer starting at rest will reach the singularity with zero  velocity as attractive and repulsive effects compensate each other along the trajectory of the observer \cite{Arrechea:2020evj}.
In this paper, we would not address the issues of validity of the $4D$ regularization procedure or an entirely consistent theory in four dimensions. We will be investigating the static spherically symmetric black hole solution, which is seemingly identical in $4D$ EGB regularized approaches.

Motivated by this, we consider the gravitational lensing by a Charged black hole in regularized $4D$ EGB gravity. Following the prescription of Bozza et al. \cite{Bozza:2001xd}, we determine the strong deflection coefficients and the resulting deflection angle, which becomes unboundedly large for smaller impact parameter values. We investigate the effect of charge on positions and magnifications of the source's relativistic images. We also obtained the corrections in the deflection angle due to the GB coupling parameter in the supermassive black hole contexts.

The rest of the paper is organized as follows. In the  Sect.~\ref{sect2}, we discuss the static spherically symmetric Charged black hole in $4D$  EGB gravity. Formalism for gravitational bending of light in strong-field limit is setup in Sect.~\ref{sect3}, whereas strong-lensing observables, numerical estimations of deflection angle, and image positions and magnifications are presented in Sect.~\ref{sect4}. Lensing by supermassive black holes Sgr A* and M87* is discussed in Sect.~\ref{sect5}.  Finally, we summarize our main findings in  Sect.~\ref{sect6}.\\

\section{Charged black holes in $4D$ EGB gravity }\label{sect2}
The EGB gravity action with re-scaled coupling constant $\alpha/(D-4)$ and minimally coupled electromagnetic field in $D$ dimensional spacetime reads \cite{Fernandes:2020rpa}
\begin{eqnarray}
S &=& \frac{1}{2} \int d^{D}x\sqrt{-g}[R + \frac{\alpha}{D-4} \mathscr{G}-F_{\mu\nu}F^{\mu\nu}],\label{action}
\end{eqnarray} 
with  $\mathscr{G}$ is the GB term defined by 
\begin{equation}
\mathscr{G}= R^2 -4 R_{\mu \nu}R^{\mu \nu}+R_{\mu \nu \rho \sigma}R^{\mu \nu \rho \sigma},
\end{equation}
 $g$ is the determinant of metric tensor $g_{\mu\nu}$, $R$ is the Ricci scalar, and $F_{\mu\nu}=\partial_{\mu} A_{\nu}-\partial_{\nu}A_{\mu}$ is the Maxwell tensor with $A_{\mu}$ being the gauge potential. On varying the action (\ref{action}) with respect to the metric tensor $g_{\mu\nu}$, we obtain the field equations
\begin{equation}\label{Einsteineqn}
G_{\mu\nu} +\frac{\alpha}{D-4}H_{\mu\nu}=T_{\mu\nu}\equiv 2\Bigr( F_{\mu\sigma}F_{\nu}{^\sigma}-\frac{1}{4}g_{\mu\nu}F_{\alpha\beta}F^{\alpha\beta} \Bigl),
\end{equation}
where $G_{\mu\nu}=R_{\mu\nu}-\frac{1}{2}Rg_{\mu\nu}$ is the Einstein's tensor, and $H_{\mu\nu} $ is the Lanczos tensor \cite {Lanczos:1938sf} and is given by
\begin{eqnarray}
H_{\mu\nu}&=&2(RR_{\mu\nu}-2R_{\mu\sigma}R_{\ \nu}^{\sigma}-2R_{\mu\sigma\nu\rho}R^{\sigma\rho}-R_{\mu\sigma\rho\beta}R_{\nu}^{\sigma\rho\beta}) \nonumber \\ 
&&- \frac{1}{2}g_{\mu\nu}\mathscr{G},
\end{eqnarray}
and $T_{\mu\nu}$ is the energy-momentum tensor for the electromagnetic field. Considering a static and spherically symmetric $D$-dimensional metric anstaz
\begin{eqnarray}
ds^2 &=& -f(r) dt^2 + \frac{1}{f(r)}  dr^2 + r^2 d\Omega_{D-2}^2,\label{NR}
\end{eqnarray}
and where $  d\Omega_{D-2}$ is the metric of a $(D-2)$-dimensional spherical surface. Solving the field equations (\ref{Einsteineqn}), in the limit $D\to 4$,  yields a solution \cite{Fernandes:2020rpa}
\begin{equation}\label{fr}
f_{\pm}(r)= 1+\frac{r^{2}}{2\alpha}\left(1\pm\sqrt{1+4\alpha\left(\frac{2M}{r^{3}}-\frac{Q^{2}}{r^{4}}\right)}\right).
\end{equation}
Here, $M$ and $Q$ can be identified, respectively, as the mass and charge parameters of the black hole. In the limit of $Q=0$, metric (\ref{NR}) with (\ref{fr}) corresponds to the static $4D$ EGB black hole \cite{Glavan:2019inb}.  The metric (\ref{NR})  is not singularity free as the curvature invariants diverge at $r=0$. A freely-falling test particle with zero angular momentum falls into the singularity within a finite proper time, indeed the effective potential identically vanishes for the radially moving photon (cf. Eq.~\ref{geoEq}). In addition, it has also been proven for uncharged $4D$ EGB black hole that central curvature singularity can be reached by radial freely-falling observers within a finite proper time \cite{Arrechea:2020evj}. The Equation~(\ref{fr}) corresponds to the two branches of solutions depending on the choice of ``$\pm$", such that at large distances Eq.~(\ref{fr}) reduces to  
\begin{align}
f_-(r)&=1-\frac{2M}{r}+\frac{Q^2}{r^2}+\mathcal{O}\left(\frac{1}{r^4}\right),\nonumber\\
f_+(r)&=1+\frac{2M}{r}-\frac{Q^2}{r^2}+\frac{r^2}{\alpha } +\mathcal{O}\left(\frac{1}{r^4}\right).
\end{align}
In the vanishing limit of $\alpha$ only -ve branch smoothly recovers the Reissner-Nordstrom black hole \cite{Fernandes:2020rpa}. Thus, we will limit our discussions for -ve branch only. The effect of GB coupling parameter faded at large distances, as the Charged black holes in $4D$ EGB gravity (\ref{fr}) smoothly retrieve the Reissner-Nordstrom black hole ($\alpha=0$). However, one can expect considerable departure in a strong-field regime where usually full features of GB corrections come in to play. The Charged black holes in $4D$ EGB gravity are characterized uniquely by mass $M$, charge $Q$, and GB parameter $\alpha$. 
\begin{figure}[b!]
	\includegraphics[scale=0.8]{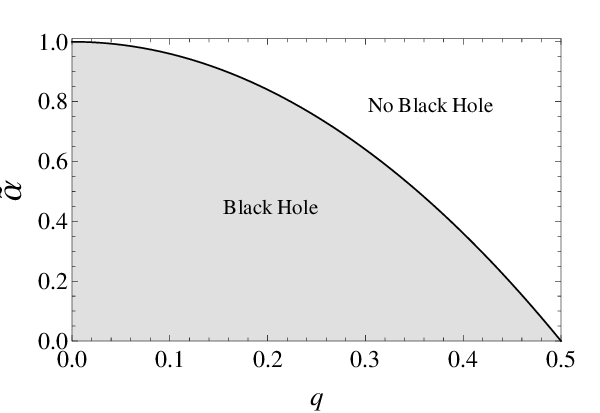}
	\caption{The parameter space ($q, \tilde{\alpha}$) for the existence of black hole horizons; black solid line corresponds to the values of parameters for which extremal black hole exists.}\label{ParReg}
\end{figure}
To begin a discussion on the strong gravitational lensing, we adimensionlise the Charged black hole metric of EGB gravity (\ref{NR}) in terms of Schwarzschild radius $2M$ by defining $x=r/2M$, $T=t/2M$, $\tilde{\alpha}=\alpha/M^2$, and $q=Q/2M$. Then we have
\begin{figure*}
\begin{center}	
\begin{tabular}{c c c c}
\includegraphics[scale=0.73]{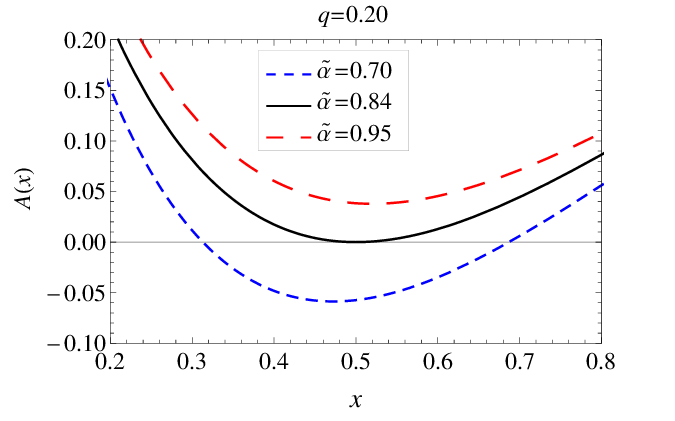}&
\includegraphics[scale=0.73]{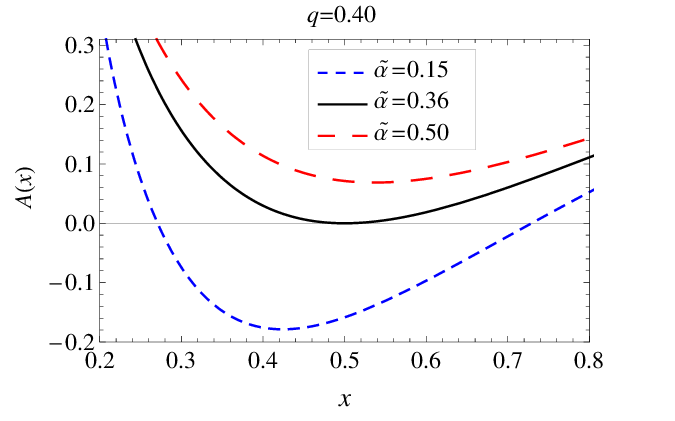}\\
\includegraphics[scale=0.73]{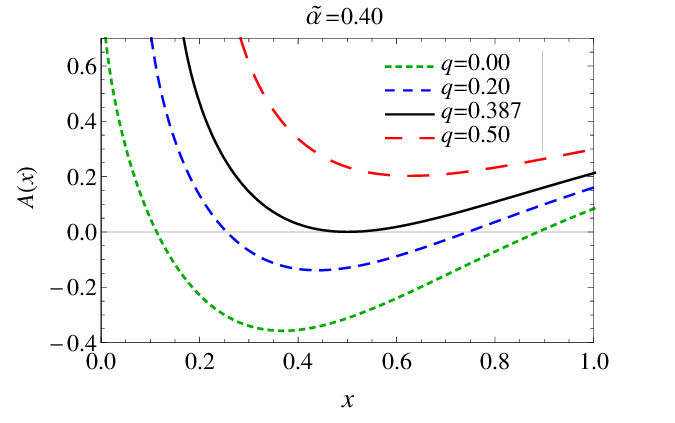}&
\includegraphics[scale=0.73]{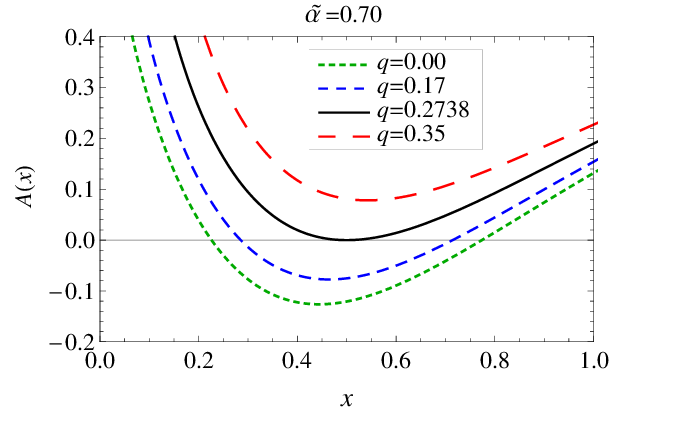}
\end{tabular}
\caption{(Upper panel) Plot showing the horizons for various values of GB coupling parameter $\tilde{\alpha}$ and fixed values of $q$. (Lower panel) Plot showing the horizons for different values of  $q$ and fixed $\tilde{\alpha}$. The black solid lines correspond to the extremal black holes. }\label{plot}	
\end{center}	
\end{figure*}
\begin{figure*}            
	\begin{tabular}{c c c c}
		\includegraphics[scale=0.7]{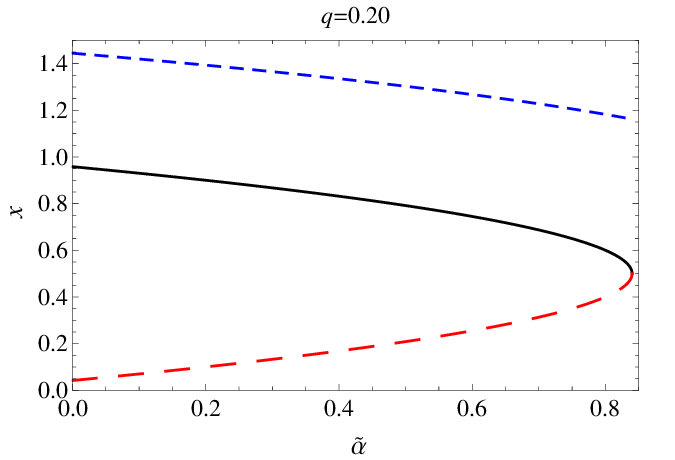}&
		\includegraphics[scale=0.7]{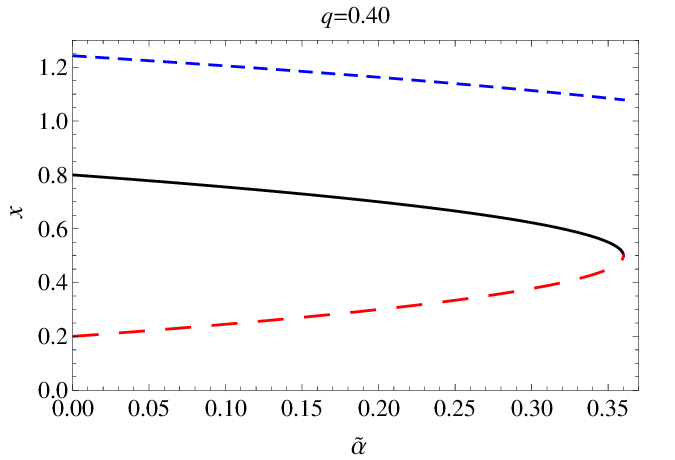}\\
		\includegraphics[scale=0.7]{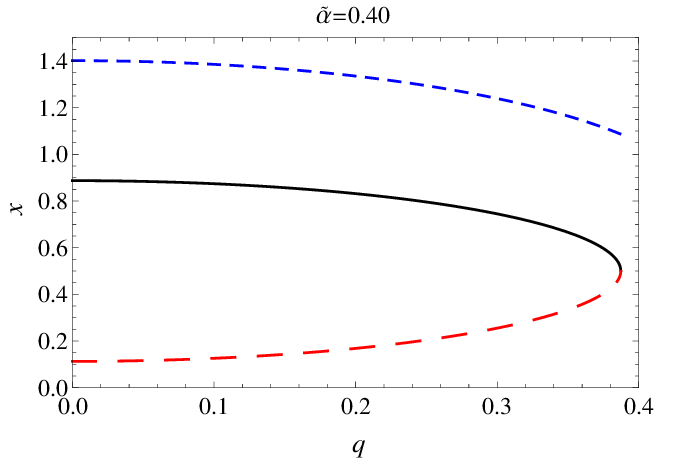}&
		\includegraphics[scale=0.7]{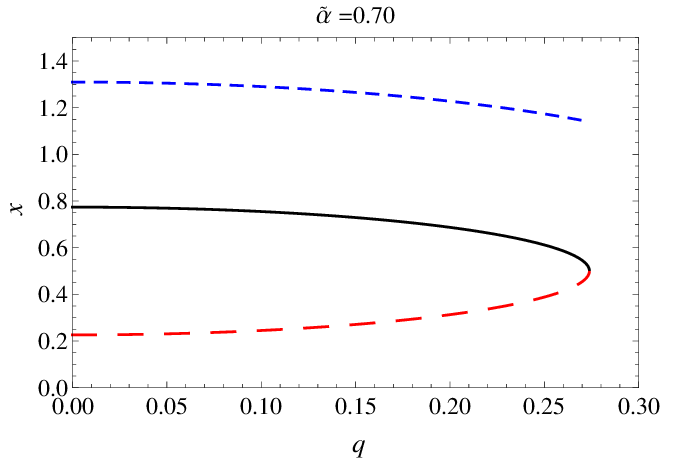}
	\end{tabular}
	\caption{The behavior of event horizon radii $x_+$ (solid black line), Cauchy horizon radii $x_-$ (dashed red line), and photon sphere radii $x_m$ (dashed blue line) with  $\tilde{\alpha}$ and $q$.   }\label{plot2}		
\end{figure*}
\begin{align}\label{NR1}
d\tilde{s}^2 &= (2M)^{-2} ds^2 = -A(x)dT^2+\frac{1}{A(x)}dx^2\nonumber\\
&+C(x)(d\theta ^2 +\sin ^2\theta\,d\phi^2),
\end{align}   
where 
\begin{equation}
A(x) = 1+\frac{2 x^{2}}{\tilde{\alpha}}\left(1\pm\sqrt{1+\tilde{\alpha}\left(\frac{1}{x^{3}}-\frac{q^{2}}{x^{4}}\right)}\right) , \quad C(x)=x^2.
\end{equation}
It is clear that metric (\ref{NR1}) possess a coordinate singularity at \begin{eqnarray}
A(x)=0\Rightarrow 4\tilde{\alpha} x^2 -4\tilde{\alpha} x +\tilde{\alpha}(\tilde{\alpha} + 4q^2)=0,
\end{eqnarray}
which admits up to two real positive roots given by
\begin{eqnarray}
x_{\pm} &=& \frac{1}{2} \left(1 \pm \sqrt{ 1- 4 q^2 - \tilde{\alpha}}\right).\label{horizon}
\end{eqnarray}
The two roots $ x_{\pm} $ correspond  to the radii of black hole event (outer) horizon ($x_+$) and Cauchy (inner) horizon ($x_-$). It is clear from Eq.~(\ref{horizon}) that for the existence of black hole, the allowed values for $\tilde{\alpha}$ are given by
\begin{eqnarray}
\tilde{\alpha} \leq 1-4q^2  & \text{for}\  0\leq q\leq 0.5.
\end{eqnarray}
The GB coupling parameter is related with the inverse string tension and hence is a positive entity. For a given value of $q$, there always exists an extremal value of $\tilde{\alpha}=\tilde{\alpha}_e=1-4q^2$, for which black hole possess  degenerate horizons, i.e., $x_-=x_+=x_e$, such that $\tilde{\alpha}<\tilde{\alpha}_e$ leads to two distinct horizons and $\tilde{\alpha}>\tilde{\alpha}_e$ leads to no-horizons (cf. Fig.~\ref{ParReg} and Fig.~\ref{plot}). Similarly, one can find the extremal value of $q_e=\sqrt{1-\tilde{\alpha}}/2$, for a given value of $\tilde{\alpha}$. In Fig.~\ref{ParReg}, the parameter space ($q, \tilde{\alpha}$) is shown, the black curve comprises the extremal values of the parameters which lead to the existence of extremal black holes. The behaviour of horizon radii with GB coupling parameter $\tilde{\alpha}$ and black hole charge $q$ is shown in Fig.~\ref{plot2}. Event horizon radius decreases whereas Cauchy horizon radius increases with increasing $q$ or $\tilde{\alpha}$, such that  Charged black holes of $4D$ EGB gravity always possess smaller event horizon as compared to Schwarzschild and Reissner-Nordstrom black holes. The bounds on the GB parameter have also been obtained in the context of gravitational instability \cite{Konoplya:2020bxa,Konoplya:2020juj,Konoplya:2020der} and with the aid of recent M87* black hole shadow observations \cite{ShaRef}.

\section{Light deflection angle}\label{sect3}
\begin{figure}[h!]
\begin{center}	
	\includegraphics[scale=0.35]{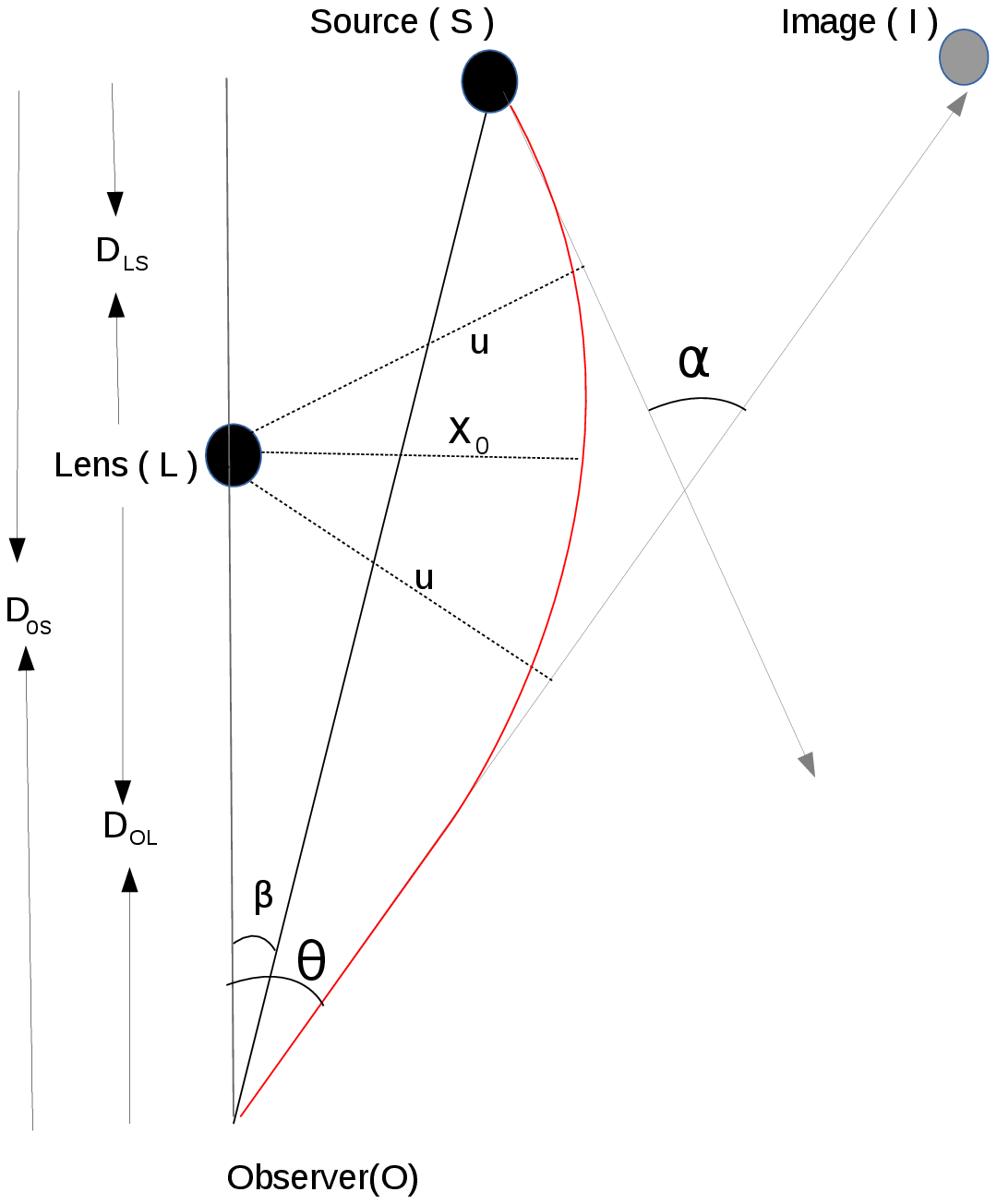}
	\caption{Schematic for geometrical configuration of gravitational lensing.}\label{fig1}		
\end{center}
\end{figure}

In this section, we investigate the strong gravitational lensing in the Charged black holes of $4D$ EGB gravity to compute the deflection angles, location of relativistic images, their magnifications and the effect of $\tilde{\alpha}$ and $q$ on them. We consider that light source $S$ and observer $O$ are sufficiently far from the black hole $L$, which acts as a lens, and they are nearly aligned. The light ray emanating from the source travel in a straight path towards the black and only when it encounters the black hole gravitational field it suffers from the deflection (cf. Fig.~\ref{fig1}). The amount of deflection suffered by light depends on the impact parameter $u$ and distance of minimum approach $x_0$, at which light suffer reflection and starts outward journey toward the observer \cite{Weinberg:1972}. Consider the propagation of light on the equatorial plane $(\theta =\pi/2)$, as due to spherical symmetry, the whole trajectory of the photon is limited on the same plane. The projection of photon four-momentum along the Killing vectors of isometries is conserved quantities, namely the energy $\mathcal{E}=-p_{\mu}\xi^{\mu}_{(t)}$ and angular momentum $\mathcal{L}=p_{\mu}\xi^{\mu}_{(\phi)}$ are constant along the geodesics, where $\xi^{\mu}_{(t)}$ and $\xi^{\mu}_{(\phi)}$ are, respectively, the Killing vectors due to time-translational and rotational invariance \cite{Chandrasekhar:1992}.

Photons follow the null geodesics of metric (\ref{NR1}), $d\tilde{s}^2=0$, which yields
\begin{equation}
\left(\frac{dx}{d\tau}\right)^2\equiv \dot{x}^2={\cal E}^2-\frac{\mathcal{L}^2A(x)}{C(x)},\label{geoEq}
\end{equation}
where $\tau$ is the affine parameter along the geodesics. Photons traversing close to the black hole, experience radial turning points $\dot{x}=0$ and follows the unstable circular orbits, whose radii $x_m$ can be obtained from 
\begin{eqnarray}
\frac{A'(x)}{A(x)} &=& \frac{C'(x)}{C(x)},
\end{eqnarray}
which reduces to 
\begin{eqnarray}\label{ps}
 4 x^4 - 9 x^2  +(24 q^2 + 4 \tilde{\alpha}) x - 4 q^2 \tilde{\alpha} -16 q^4 &=& 0
\end{eqnarray}
Here prime corresponds to the derivative with respect to the $x$ and  the eq.(\ref{ps}) admits at least one positive solution and then the largest real root  is defined as the radius of the unstable circular photon orbits (cf.~Fig. \ref{plot2}). A small radial perturbations drive these photons into the black hole or toward spatial infinity \cite{Chandrasekhar:1992}. Due to spherical symmetry, these orbits generate a photon sphere around the black hole. The radii of photon orbits for Charged black holes of $4D$ EGB gravity decrease with increasing $q$ and $\tilde{\alpha}$ (cf.~Fig. \ref{plot2}).
Further, at the distance of minimum approach, we have \cite{Weinberg:1972}
\begin{equation}
\frac{dx}{d\phi}=0,\;\;\Rightarrow\;\; u\equiv \frac{\cal L}{\cal E} =\sqrt{\frac{C(x_0)}{A(x_0)}}.
\end{equation}
Following the method developed by Bozza  \cite{Bozza:2002zj}, 
the total deflection angle suffered by the light in its journey from source to observer is given by
\begin{equation}
\alpha_D (x_0)= I(x_0)-\pi,\label{def}
\end{equation}
where 
\begin{equation}\label{def3}
I(x_0)=2\int_{x_0}^\infty \frac{d\phi}{dx}dx
=\int_{x_0}^\infty \frac{2\, dx}{\sqrt{A(x)C(x)}\sqrt{\frac{C(x)A(x_0)}{C(x_0)A(x)}-1}},
\end{equation}
The deflection angle increases as distance of minimum approach $x_0$ decreases and shows divergence as it approaches the photon sphere $x_m$ \cite{Bozza:2002zj}. In the strong deflection limit, we can expand the deflection angle near the photon sphere, for the purpose we define a new variable $z$
as \cite{Bozza:2002zj}
\begin{eqnarray}
z &=& \frac{A(x)-A(x_0)}{1-A(x_0)},
\end{eqnarray} 
the integral (\ref{def}) can be re-written as 
\begin{eqnarray}
I(x_0) &=& \int_{0}^{1}R(z,x_0) f(z,x_0) dz,\label{def1}
\end{eqnarray}
with the functions
\begin{eqnarray}
R(z,x_0) &=&  \frac{2\sqrt{C(x_0)}(1-A(x_0))}{C(x)A'(x)},\\
f(z,x_0) &=& \frac{1}{\sqrt{A(x_0)-\frac{A(x)}{C(x)}C(x_0)}}\label{f(x)},
\end{eqnarray}
where $x=A^{-1}[(1-A(x_0))z+A(x_0)]$. 
Making a Taylor series expansion of the function in Eq.~(\ref{f(x)}), we get 
\begin{eqnarray}
f_0(z,x_0) &=& \frac{1}{\sqrt{\phi(x_0) z + \gamma(x_0) z^2}} 
\end{eqnarray}
where 
\begin{align}
\phi(x_0) = &\frac{1-A(x_0)}{A'(x_0) C(x_0)}\left[C^\prime(x_0) A(x_0)-
 A^\prime(x_0) C(x_0)\right],\\
\gamma(x_0)=& \frac{\left(1-A(x_0)\right)^2}{2 A^\prime(x_0){^3} C(x_0)^2}\Big[2C(x_0)C^\prime(x_0) A^\prime(x_0){^2}\nonumber\\
&+A(x_0)A^\prime(x_0) C(x_0)C^{\prime\prime}(x_0)   -C(x_0)C^\prime(x_0)\nonumber\\
& A(x_0) A^{\prime\prime}(x_0) -2C^\prime(x_0){^2}A(x_0)A^\prime(x_0) \Big].
\end{align}
The integrand term $f(z,x_0)$ diverges for $x_0\to x_m$ leading to diverging deflection angle in Eq.~(\ref{def1}). Therefore we made Taylor series expansion to identify this diverging term and order of divergence, so that we can subtract this term from $I(x_0)$ in Eq.~(\ref{def1}) to get the regular term $I_R(x_0)$ in the strong-field regime. $R(z,x_0)$ is regular for all values of $z$, however, for $x_0=x_m$, we have $\phi(x_0)=0$ and $f_0 \approx 1/z$, which diverge as $z\to 0$. Following the above definitions, the diverging part in the integral Eq.~(\ref{def1}) can be identified as \cite{Bozza:2002zj}
\begin{equation}
I_D(x_0)=\int_0 ^1 R(0,x_m)f_0(z,x_0)dz,\label{divergent}
\end{equation}
whereas the regular part $I_R(x_0)$ is 
\begin{align}
I_R(x_0)&=I(x_0)-I_D(x_0)\nonumber\\
&=\int_0 ^1 \Big(R(z,x_0)f(z,x_0)-R(0,x_m)f_0(z,x_0)\Big)dz.\label{regular}
\end{align}
such that $I_D(x_0)$ has logarithmic divergence and $I_R(x_0)$ is regular with divergence subtracted from the complete integral (\ref{def1}). The deflection angle can be written in terms of $x_0$ as \cite{Bozza:2002zj}
\begin{eqnarray}\label{deff1}
\alpha_D(x_0) &=& -a \log\Big(\frac{x_0}{x_m}-1\Big)+ b + O(x_0-x_m),
\end{eqnarray} 
where 
\begin{eqnarray}\label{ab}
a&=& \frac{R(0,x_m) }{\sqrt{\gamma(x_m)}}, \\ 
b&=& -\pi + I_R(x_m) + a \log \left( \frac{2(1-A(x_m)}{A'(x_m). x_m}\right),
\end{eqnarray}
\begin{figure*}
	\begin{tabular}{c c}
		\includegraphics[scale=0.73]{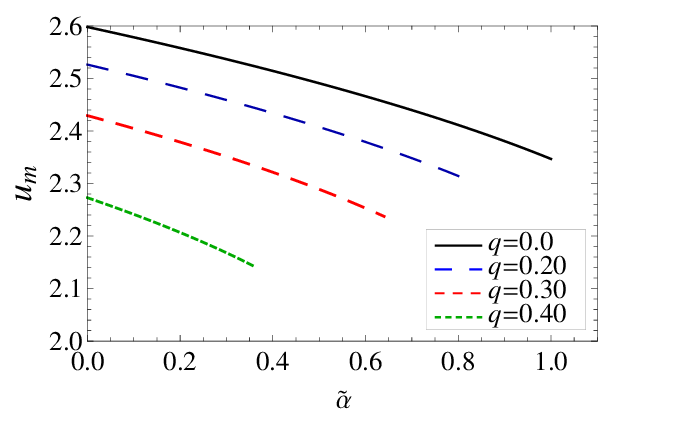}
		\includegraphics[scale=0.73]{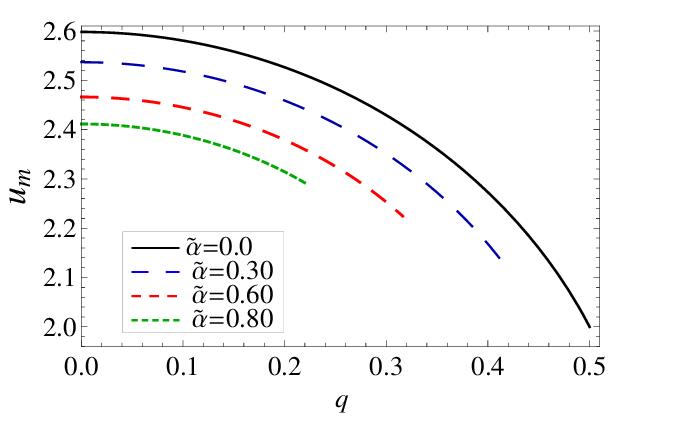}	
	\end{tabular}\caption{Plot showing the impact parameter $u_m$ for photon circular orbits with $q$ and $\tilde{\alpha}$.}\label{imp}
\end{figure*}
\begin{center}
\begin{figure*}
	\begin{centering}
		\begin{tabular}{c c c c}
			\includegraphics[scale=0.75]{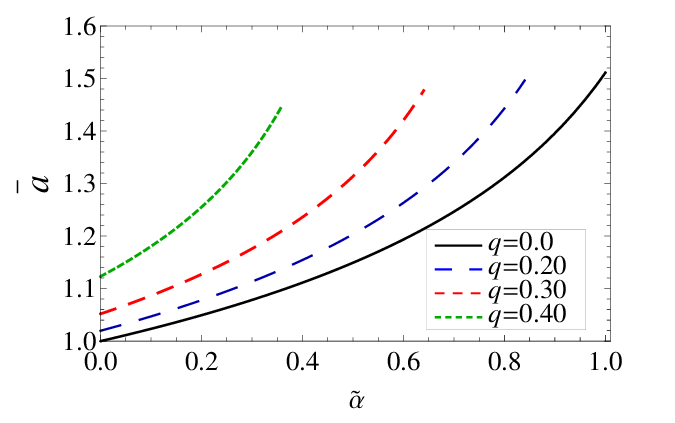}&
			\includegraphics[scale=0.75]{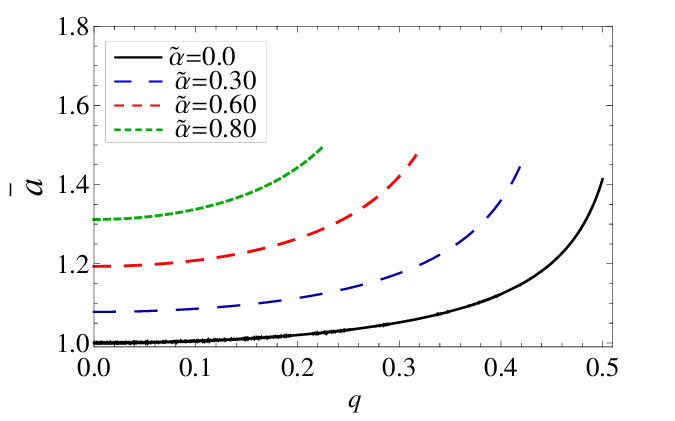}\\
			\includegraphics[scale=0.75]{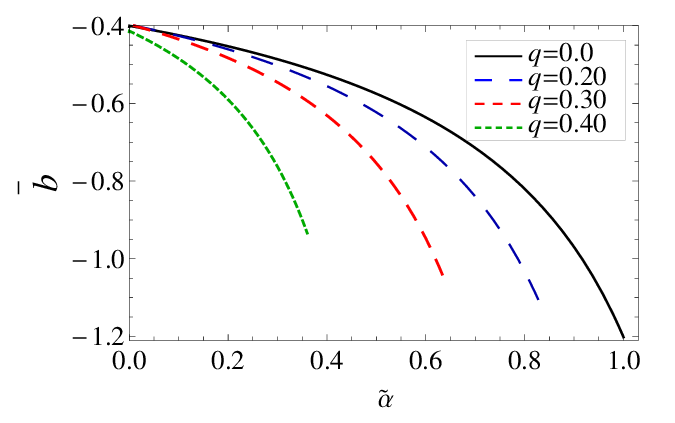}&
			\includegraphics[scale=0.75]{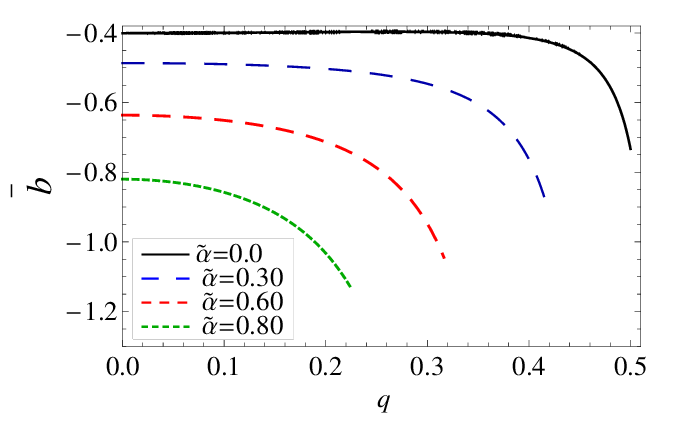}
		\end{tabular}
	\end{centering}
	\caption{Plot showing the behavior of strong lensing coefficients $\bar{a}$ and $\bar{b}$ as a functions of $\tilde{\alpha}$ (Left Panel) and $q$ (Right Panel)}\label{plot6}		
\end{figure*}
\begin{figure*}
	\begin{center}	
		\begin{tabular}{c c}
			\includegraphics[scale=0.7]{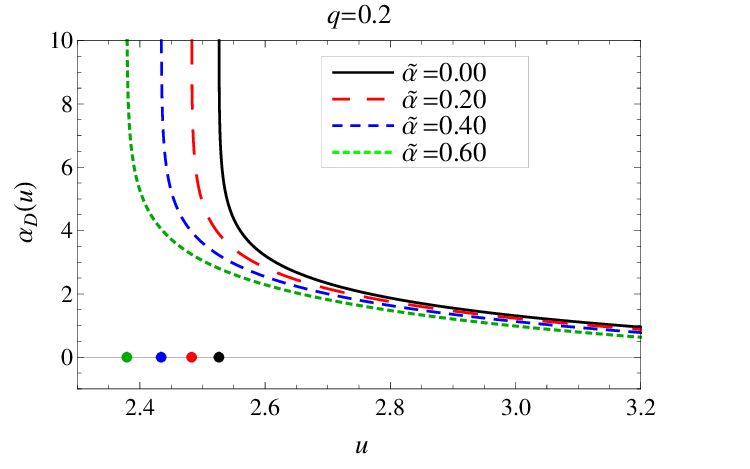}&
			\includegraphics[scale=0.7]{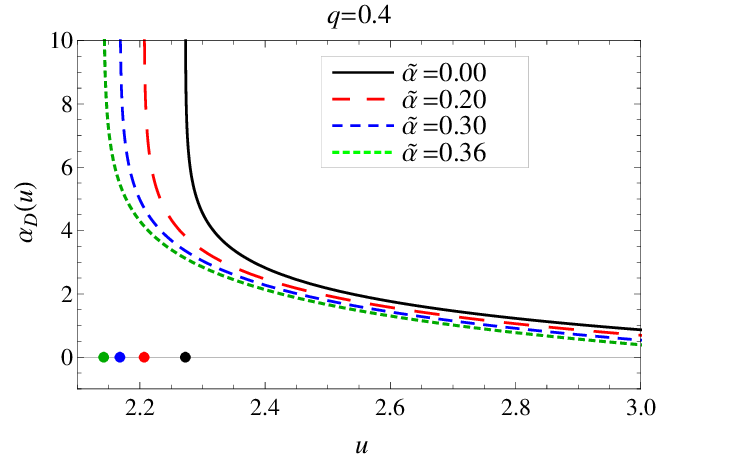}\\
			\includegraphics[scale=0.7]{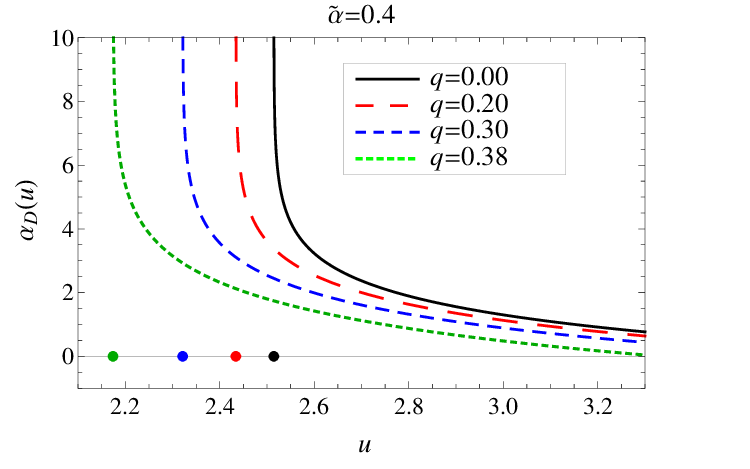}&
			\includegraphics[scale=0.7]{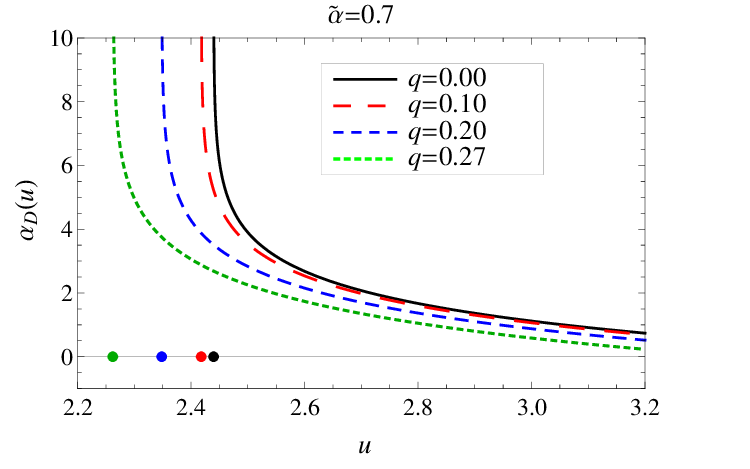}
		\end{tabular}
		\caption{Plot showing the variation of deflection as a functions impact parameter $u$ for different values of  $\tilde{\alpha}$ and $q$. The points on the horizontal axis correspond to the impact parameter $u=u_m$ at which deflection angle diverges. }\label{plot4}	
	\end{center}	
\end{figure*}
\end{center}
The Eq.~(\ref{deff1}) is coordinate dependent, however, it can be written in terms of coordinate independent variable impact parameter $u$, as follows
\begin{eqnarray}\label{defangle}
\alpha_D(u) &=& -\bar{a} \log\Big(\frac{u}{u_m}-1\Big)+\bar{b} + O(u-u_m),
\end{eqnarray} 
where
\begin{equation}
\bar{a}=\frac{a}{2}, \qquad
\bar{b}= -\pi + I_R(x_m) + \bar{a} \log \Big( \frac{2\gamma(x_m)}{A(x_m)}\Big),
\end{equation} 
We obtain the strong deflection coefficients as
\begin{align}
\bar{a} &=\frac{4\left((x_m-q^2)\tilde{\alpha} -x_m^4(c_1-1)\right)}{(c_1-1)(-\tilde{\alpha}+4x_m^3(c_1-1)) P_m},\nonumber\\
\bar{b}&= -\pi + I_R(x_m) + \bar{a} \log \left( 2\delta\right),
\end{align}
with
\begin{align}
\delta =&\frac{(c_1-1)^2}{4 x_m^2 A(x_m)}P_m^2,\\
P_m=&\frac{1}{\sqrt{c_1^3\left(\tilde{\alpha} -2x_m^2 (c_1-1)\right)}}\Big(4 x_m^6 c_1 + x_m^2 \tilde{\alpha} (-9 + 4 x_m c_1)\nonumber\\
&- 4 q^2 x_m (2 x_m^3 - 4 \tilde{\alpha} + x_m \tilde{\alpha}c_1) -8 q^4 \tilde{\alpha}\Big)^{1/2},\\
c_1 =& \sqrt{1+ \frac{(x_m-q^2) \tilde{\alpha}}{x_m^4}},
\end{align}
where $\bar{a}$ and $\bar{b}$ are called the strong deflection limit coefficients. In Fig.~\ref{imp}, we plotted the impact parameter $u_m$ for the photons moving on the unstable circular orbits around black hole as a functions of charge parameter $q$ and GB coupling parameter $\tilde{\alpha}$. It is clear that $u_m$ decrease with $q$ and $\tilde{\alpha}$. The behavior of lensing coefficients are shown in Fig.~\ref{plot6}, which in the limits of $\tilde{\alpha}\to 0$ and $q=0$, smoothly retain the values for the Schwarzschild black hole, viz., $\bar{a}=1$ and $\bar{b}=-0.4002$ \cite{Bozza:2001xd,Bozza:2002zj}. Coefficient $\bar{a}$ increases whereas $\bar{b}$ decreases with increasing $q$ or $\alpha$. The resulting deflection angle $\alpha_D(u)$ is shown as a function of impact parameter $u$ for various values of $q$ and $\tilde{\alpha}$ in Fig.~\ref{plot4}. For a fixed value of $u$, deflection angle decreases with increasing $q$ or $\tilde{\alpha}$, therefore, the deflection angle is higher for Schwarzschild and Reissner-Nordstrom black holes than those for the Charged black holes of $4D$ EGB gravity. Figure \ref{plot4} infers that the $\alpha_D(u)$ increases as impact parameter $u$ approaches the $u_m$ and becomes unboundedly large for $u=u_m$.

\section{Strong lensing observables}\label{sect4}
The deflection angle obtained in Eq.~(\ref{defangle}) is directly related to the positions and magnification of the relativistic images, which is given by lens equation \cite{Bozza:2007gt}
\begin{eqnarray}
D_{OS}\tan\beta &=& \frac{D_{OL}\sin\theta - D_{LS}\sin(\alpha-\theta)}{\cos(\alpha-\theta)}],\label{lensEq}
\end{eqnarray}
where $\theta$ and $\beta$, respectively, are the angular separations of image and source from the black hole as shown in Fig.~\ref{fig1}. The $D_{LS}$ is the distance between the source and black hole and  the distances from the observer to the source and black hole are respectively $D_{OS}$ and $D_{OL}$; all distances are expressed in terms of the Schwarzschild radius $x_s=R_s/2M$. For nearly perfect alignment of source, black hole and observer, viz. small values of $\theta$ and $\beta$, Eq.~(\ref{lensEq}) reduces to the following form \cite{Bozza:2007gt}
\begin{eqnarray}\label{beta}
\beta &=& \theta -\frac{D_{LS}}{D_{OS}}\Delta \alpha_n.
\end{eqnarray}
\begin{figure}
\begin{center}	
	\includegraphics[scale=0.75]{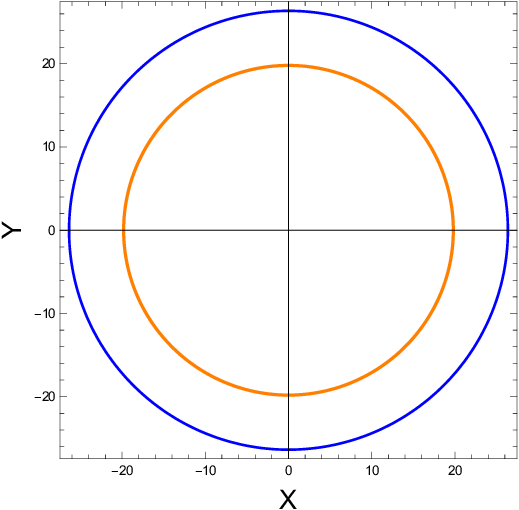}
	\caption{The outermost relativistic Einstein ring as seen in the observers sky. The blue ring corresponds for Sgr A* and orange for the M87*.}\label{fig8}		
\end{center}
\end{figure}
In case of strong lensing, photons complete multiple circular orbits around black hole before escaping toward observer, therefore $\alpha_D$ can be replaced by $2n\pi + \Delta \alpha_n$ in Eq.~(\ref{lensEq}), with $n \in N$ and $0 < \Delta \alpha_n \ll 1$. \\
Now, making a Taylor series expansion of the deflection angle about $(\theta_n ^0)$ to the first order, as
\begin{equation}\label{Taylor}
\alpha_D(\theta) = \alpha_D(\theta_n ^0) +\frac{\partial \alpha_D(\theta)}{\partial \theta } \Bigg |_{\theta_n ^0}(\theta-\theta_n ^0)+O(\theta-\theta_n ^0),
\end{equation}
Using the deflection angle Eq.~(\ref{defangle}), Eq.~(\ref{Taylor}) reduces to 
\begin{eqnarray}
\Delta\alpha_n &=& -\frac{\bar{a}D_{OL}}{u_m e_n}\Delta\theta_n.
\end{eqnarray}
The angular separation between the lens and $n$th relativistic images now can be written as 
\begin{eqnarray}
\theta_n\simeq \theta_n ^0 + \Delta\theta_n,\label{angsep}
\end{eqnarray}
where 
\begin{eqnarray}
\theta_n ^0 &=& \frac{u_m}{D_{OL}}(1+e_n),\nonumber\\
\Delta\theta_n &=& \frac{D_{OS}}{D_{LS}}\frac{u_me_n}{D_{OL}\bar{a}}(\beta-\theta_n ^0),\nonumber\\
e_n &=& e^{{\bar{b}-2n\pi}/{\bar{a}}}, \label{obs}
\end{eqnarray}
\begin{center}
\begin{figure*}[t]
	\begin{tabular}{c c}
		\includegraphics[scale=0.75]{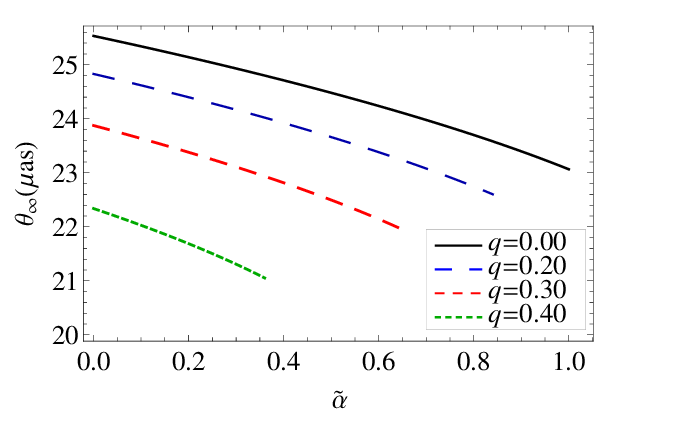}&
		\includegraphics[scale=0.75]{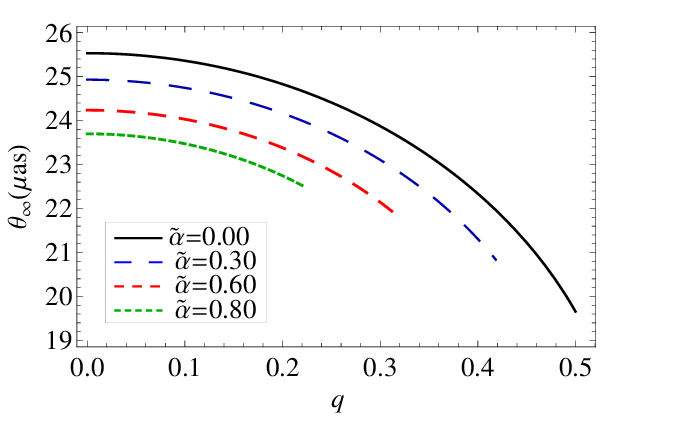}\\
		\includegraphics[scale=0.75]{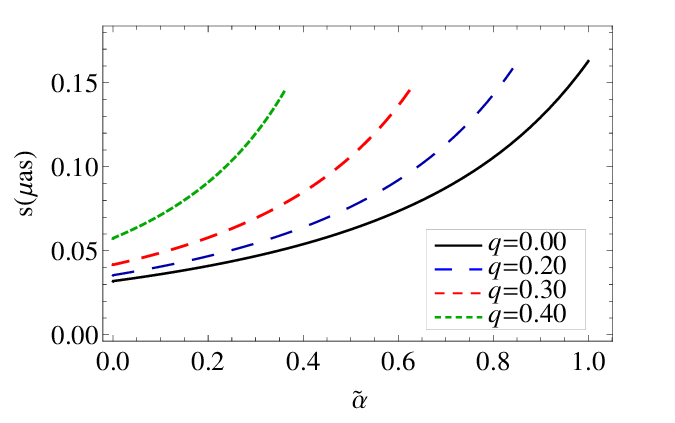}&
		\includegraphics[scale=0.75]{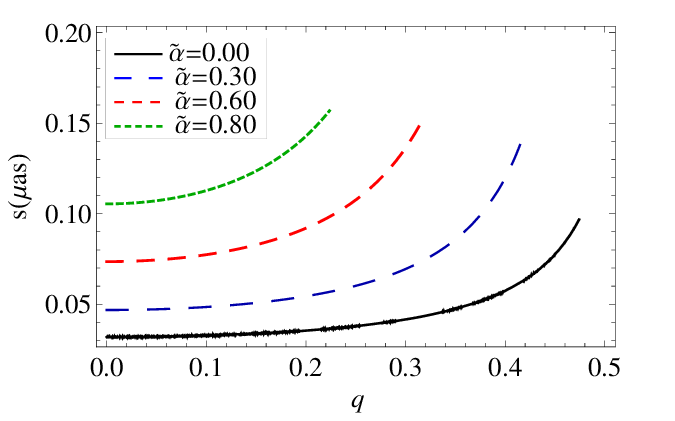}\\
		\includegraphics[scale=0.75]{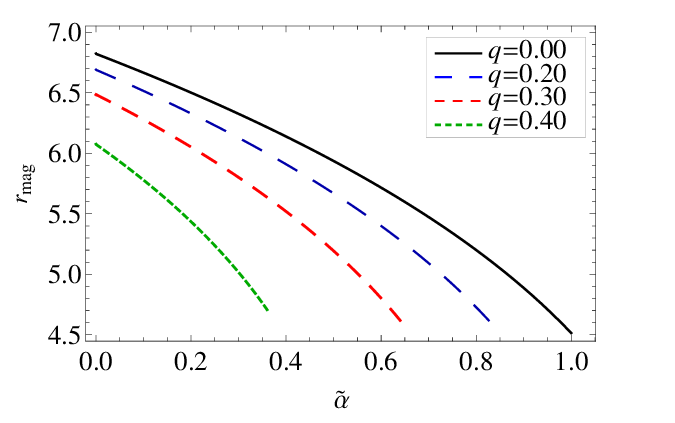}&
		\includegraphics[scale=0.75]{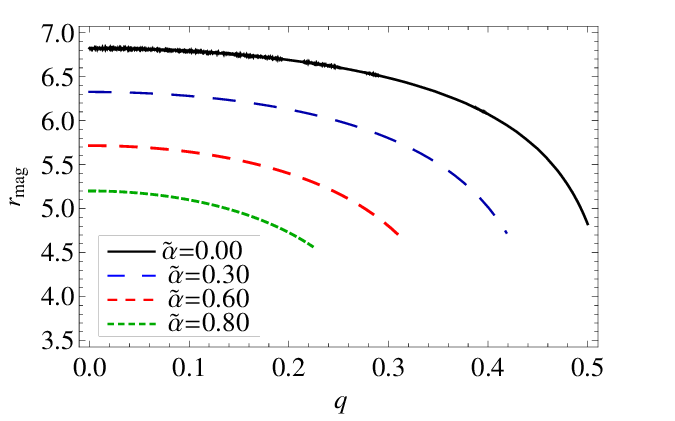}
	\end{tabular}
	\caption{Plots showing the behavior of strong lensing observables $\theta_\infty$, $s$, and $r_{\text{mag}}$ as a function $\tilde{\alpha}$ (Left panel) and  $q$ (Right panel) for Sgr A* black hole.}\label{plot7}		
\end{figure*} 
\begin{figure*}[t]
	\begin{tabular}{c c}
		\includegraphics[scale=0.75]{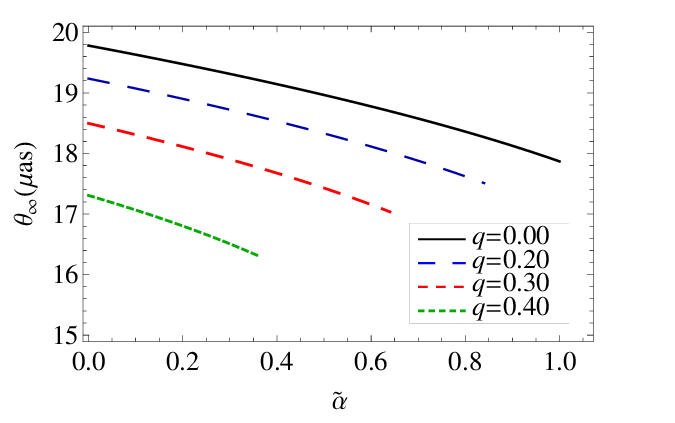}&
		\includegraphics[scale=0.75]{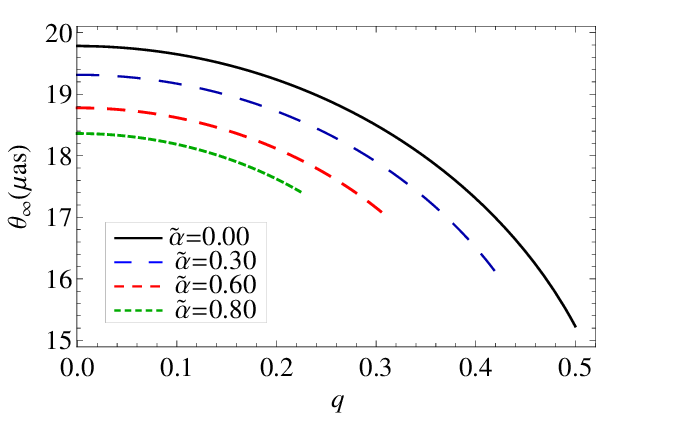}\\
		\includegraphics[scale=0.75]{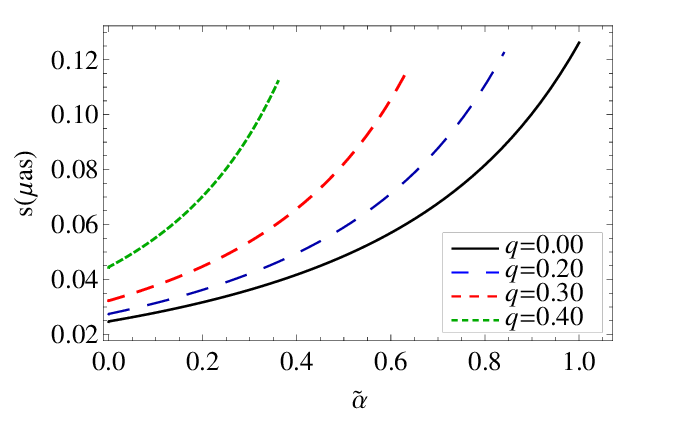}&
		\includegraphics[scale=0.75]{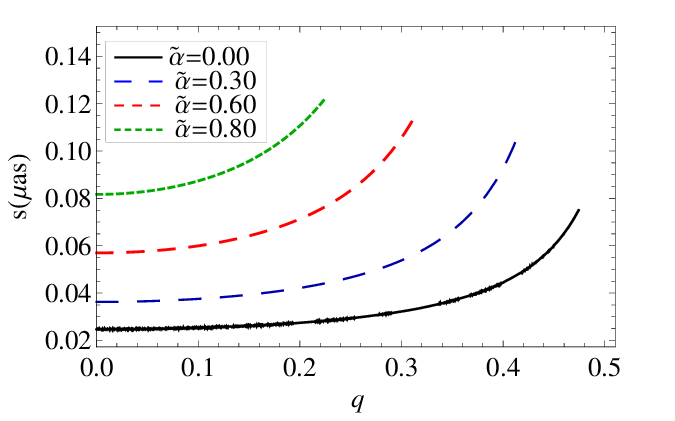}\\
		\includegraphics[scale=0.75]{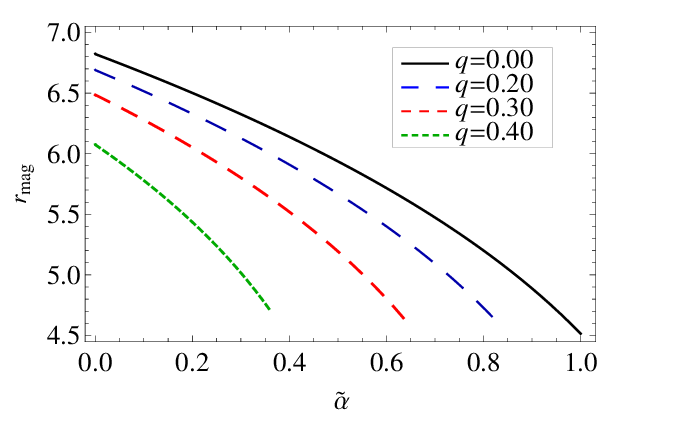}&
		\includegraphics[scale=0.75]{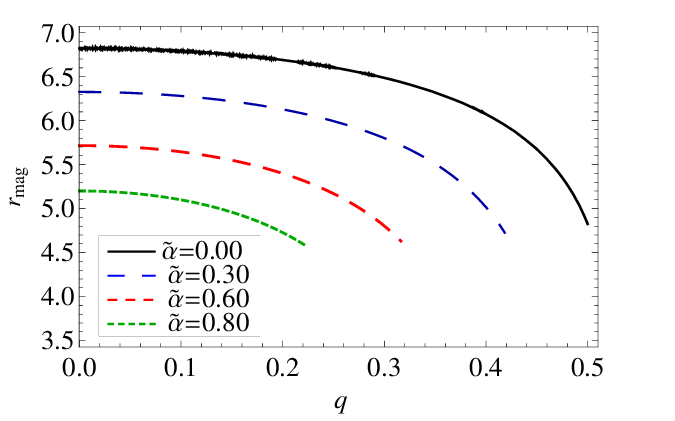}
	\end{tabular}
	\caption{Plots showing the behavior of strong lensing observables $\theta_\infty$, $s$, and $r_{\text{mag}}$ as a function $\tilde{\alpha}$ (Left panel) and  $q$ (Right panel) for M87* black hole.}\label{plot8}		
\end{figure*} 
\end{center}
where $\theta_n ^0$ corresponds to the angular separation when photon winds completely $2n\pi$  around the black hole and $\Delta\theta_n$ corresponds to the  part exceeding $2n\pi$.
\subsection{Einstein ring}
It was demonstrated that a source in front of the lens could yield relativistic images and Einstein rings \cite{Luminet} or gravitational field gives rise to an Einstein ring when the source, lens and observer are perfectly aligned. However, it is sufficient that just one point of the source is perfectly aligned to build a full relativistic Einstein ring \cite{Bozza:2004kq}. Thus for the particular configuration of source, lens and the observer being aligned (such that $\beta=0$), the Eq.~(\ref{angsep}) reduces to
\begin{eqnarray}\label{Ering}
\theta_n^{E} &=& \left(1-\frac{D_{OS}}{D_{LS}}\frac{u_me_n}{D_{OL}\bar{a}} \right) \theta_n^{0},
\end{eqnarray} 
and give radii of the Einstein rings. For a particular case when the lens is at exactly midway between the source and the observer, Eq.~(\ref{Ering}) yields
\begin{eqnarray}\label{Ering2}
\theta_n^{E} &=& \left(1-\frac{2 u_me_n}{D_{OL}\bar{a}} \right) \left(\frac{u_m}{D_{OL}}(1+e_n)\right).
\end{eqnarray} 
Since $D_{OL} \gg u_m$, the Eq.~(\ref{Ering2}) yields 
\begin{eqnarray}\label{Ering3}
\theta_n^{E} &=& \frac{u_m}{D_{OL}}\left(1+e_n \right),
\end{eqnarray} 
which gives the radius of the $n$th relativistic Einstein ring. Note that $n=1$ represents the outermost ring, and as $n$ increases, the  radius of the ring decreases. Also, it can be conveniently determined from Eq.~(\ref{Ering3}) that radius of the Einstein ring increases with the mass of the black hole and decreases as the distance between the observer and lens increases. In the Fig.~\ref{fig8},  we depict the outermost Einstein rings for Sgr A* and M87* black holes.

Similarly, magnification of images is another good source of information, which is defined as the ratio between the solid angles subtended by the image and the source, and for small angles, it is given by \cite{Bozza:2002zj}
\begin{eqnarray}\label{magn}
\mu_n &=& \left(\frac{\beta}{\theta} \;  \;\frac{d\beta}{d\theta} \Bigg|_{\theta_n ^0}\right)^{-1}.
\end{eqnarray}  
Using Eqs.~(\ref{beta}) and (\ref{obs}), the magnification (\ref{magn}) becomes:
\begin{equation}
\mu_n =\frac{1}{\beta}\Bigg( \theta_n{^0} + \frac{D_{OS}}{D_{LS}}\frac{u_me_n}{D_{OL}\bar{a}}(\beta-\theta_n{^0})\Bigg) \Bigg(\frac{D_{OS}}{D_{LS}}\frac{u_me_n}{D_{OL}\bar{a}}\Bigg),
\end{equation}
which can be simplified further by making Taylor series expansion, thus the magnification of $n$th image on both the sides is given by  \cite{Bozza:2002zj}
\begin{eqnarray}
\mu_n &=& \frac{1}{\beta} \Bigg[\frac{u_m}{D_{OL}}(1+e_n) \Bigg(\frac{D_{OS}}{D_{LS}}\frac{u_me_n}{D_{OL}\bar{a}}  \Bigg)\Bigg].
\end{eqnarray}
Thus magnification decreases exponentially  with winding number $n$ and the higher-order images become fainter. In order to  relate the results obtained analytically with observations, Bozza defined the following observables \cite{Bozza:2002zj}
\begin{eqnarray}
\theta_\infty &=& \frac{u_m}{D_{OL}},\\
s &=& \theta_1-\theta_\infty,\\
r_{\text{mag}} &=& \frac{\mu_1}{\sum{_{n=2}^\infty}\mu_n }.
\end{eqnarray} 
where $u_m$ can be written as 
\begin{eqnarray}
u_m &=& \frac{x_m}{\sqrt{1-\frac{2x_m^2}{\tilde{\alpha}}(c_1-1)}}.
\end{eqnarray}

\begin{table*}[t]
	\resizebox{1\textwidth}{!}{
		\begin{centering}	
			\begin{tabular}{p{1.2cm} p{1.5cm} p{1.5cm} p{1.5cm} p{2.5cm} p{1.5cm} p{1.5cm} p{2.5cm} p{1.5cm} p{1.5cm} p{1.5cm} }
			\hline\hline
				\multicolumn{2}{c}{Parameters}&
				\multicolumn{2}{c}{Sgr A*} & \multicolumn{4}{c}{M87*}  & \multicolumn{2}{c}{Lensing Coefficients}\\
				{$\tilde{\alpha}$ } & {$q$} & {$\theta_\infty $ ($\mu$as)} & {$s$ ($\mu$as) } &  {$r_{\text{mag}}$} & {$\theta_\infty $ ($\mu$as)} & {$s$ ($\mu$as) } &  {$r_{\text{mag}}$} & {$\bar{a}$}&{$\bar{b}$} & {$u_m/R_s$}\\\hline
				\hline
				0.00   & 0.00  & 26.327 & 0.0329  & 6.825  & 19.782 & 0.02468  & 6.825  & 1.000  & -0.4004 & 2.597  \\      
				0.00   & 0.10  & 26.151 & 0.0338  & 6.792  & 19.649 & 0.02527  & 6.792  & 1.004 & -0.4004 & 2.580  \\      
				0.00   & 0.20  & 25.602 & 0.0366 & 6.689  & 19.236 & 0.02743  & 6.689  & 1.019  & -0.4004 & 2.526  \\      
				0.00   & 0.30  & 24.617 &0.0430  & 6.483  & 18.497 & 0.03234  & 6.483  & 1.052  & -0.4004 & 2.429  \\      
				0.00   & 0.40  & 23.032 & 0.0594  & 6.072  & 17.306 & 0.04456  &6.072  & 1.123  & -0.4142 & 2.272  \\      
				0.00   & 0.50  & 20.263 & 0.1425  & 4.823  & 15.228 & 0.10655  & 4.823  &1.414  & -0.7375 & 1.999  \\      
				\hline 
				0.40   & 0.00  & 25.482 & 0.0555 & 6.138 & 19.145 & 0.0417 & 6.138 & 1.111  & -0.5263 & 2.514 \\                                     
				0.40	& 0.20  & 24.667 & 0.0659 & 5.909 & 18.533 & 0.0495 & 5.909 & 1.154  & -0.555 & 2.434 \\      
				0.40	& 0.30  & 23.527 & 0.0875 & 5.519 & 17.676 & 0.0657 & 5.519 & 1.236  & -0.6310 & 2.321 \\      
				0.40	& 0.38 & 22.032 & 0.1405 & 4.805 & 16.553 & 0.1056 & 4.805 & 1.419  & -0.8926 & 2.174\\                                  
				\hline
		    	0.70 & 0.00 & 24.725 & 0.0901 & 5.473 & 18.576 & 0.0677 & 5.473 & 1.246  & -0.7145 & 2.439 \\
			    0.70	& 0.10 & 24.504 & 0.0955 & 5.389 & 18.410 & 0.0717 & 5.389 & 1.265 & -0.7380 & 2.417 \\
				0.70 & 0.20 & 23.798 & 0.1168 & 5.092 & 17.880 & 0.0877 & 5.092 & 1.339  & -0.8395 & 2.348 \\
				0.70 & 0.27 & 22.419 & 0.1553 & 4.637 & 17.227 & 0.1167 & 4.637 & 1.471  & -1.0638 & 2.262 \\
				\hline\hline
			\end{tabular}
		\end{centering}	}	
	\caption{Estimates for lensing observables and strong lensing coefficients for the supermassive black holes Sgr A* an M87* for different values of $q$ and $\tilde{\alpha}$. $R_s = 2GM/c^2$ is the Schwarzschild radius.  }
		\label{table3}  
\end{table*}  
Since the outermost relativistic image is the brightest, the quantity $s$ is the angular separation between the outermost image from the remaining bunch of relativistic images, $r_{\text{mag}}$ is the ratio of the received
flux between the first image and all the others images clustered at $\theta_{\infty}$. For a nearly perfect alignment of source, black hole and observer, these observables can be simplified to \cite{Bozza:2002zj}
\begin{eqnarray}
s &=& \theta_\infty (e^{\frac{\bar{b}-2\pi}{\bar{a}}}),\\
r &=& e^{\frac{2 \pi}{\bar{a}}}.
\end{eqnarray}
Once these strong lensing observables are known from observations, one can estimate the strong lensing coefficients $\bar{a}$ and $\bar{b}$ and compare with the theoretically calculated values. 

\section{Lensing by Supermassive black holes}\label{sect5}	
For numerical estimation of strong lensing observables, we consider realistic cases of supermassive black holes Sgr A* and M87*, respectively, at the center of our galaxy Milky Way and nearby galaxy M87. Taking their masses $M$ and distances from earth $D_{OL}$ as, $M=4.3 \times 10^6 M_{\odot}$ and $D_{OL}=8.35\times 10^3$ pc for Sgr A* \cite{Do:2019txf}, and $M=(6.5\pm 0.7)\times 10^9 M_{\odot}$ and $D_{OL}=(16.8\pm 0.8)$ Mpc for M87* \cite{Akiyama:2019cqa}, we have tabulated the observables for various values of $q$ and $\tilde{\alpha}$ in Table (\ref{table3}). We compared the results with those for the Schwarzschild ($\tilde{\alpha}=0, q=0$) and Reissner-Nordstrom black holes ($\tilde{\alpha}=0$). It is worth to notice that, for Charged black holes of $4D$ EGB gravity, the angular separation between images are higher whereas their magnification are lower than those for the Schwarzschild and Reissner-Nordstrom black holes. Furthermore, for fixed values of parameters, Sgr A* black hole causes the larger angular separation between relativistic images than the M87* black hole, e.g., for $\tilde{\alpha}=0.40$ and $q=0.20$, $s=0.0639\mu$as for Sgr A* black hole and $s=0.0495\mu$as for M87* black hole. Charged black holes of $4D$ EGB gravity cause the larger separation $s$ and smaller magnification $r_{\text{mag}}$ as compared to the uncharged black hole of $4D$ EGB gravity ($q=0$) (cf. Table \ref{table3}). Lensing coefficients $\bar{a}$ increases whereas $\bar{b}$ decreases with increasing $q$ or $\tilde{\alpha}$ (cf. Table \ref{table3} and Fig.~\ref{plot6}).

We plotted lensing observables $\theta_{\infty}, s$, and $r_{\text{mag}}$ against $\tilde{\alpha}$ and $q$ for the Sgr A* and M87* black holes, respectively, in Fig.~(\ref{plot7}) and Fig.~(\ref{plot8}). For a given value of $D_{OL}$, the limiting value of angular position $\theta_{\infty}$ is smaller than Schwarzschild case and decreases with $q$ or $\tilde{\alpha}$. On the other hand the separation between the images $s$, for the Charged black holes of $4D$ EGB gravity is larger than the Schwarzschild black hole and it further increases with $q$ or $\tilde{\alpha}$. However, the relative magnification $r_{\text{mag}}$ decrease with increasing  $q$ or $\tilde{\alpha}$. So images are far away from the black hole and thereby less packed than in the Schwarzschild case. 

\section{Conclusion}\label{sect6}
The regularized $4D$ EGB gravity  proposed in  \cite{Tomozawa:2011gp,Cognola:2013fva,Glavan:2019inb} is  characterized by the non-trivial contribution of the GB quadratic term to the gravitational dynamics in $4D$ spacetime. Thereby, this  $4D$ EGB gravity with quadratic-curvatures bypasses the Lovelock's theorem and yields the diffeomorphism invariance and second-order equations of motion. It is shown by Glavan and Lin \cite{Glavan:2019inb} that the $4D$ EGB gravity possesses only the degrees of freedom of a massless graviton and thus free from the instabilities. Further, static and spherically symmetric black hole solutions of this $4D$ EGB gravity are also valid in other theories of gravity \cite{Cai:2009ua,Casalino:2020kbt,Lu:2020iav,Hennigar:2020lsl}. 

With this motivation, we have analysed the strong gravitational lensing of light due to static spherically symmetric Charged black holes to $4D$ EGB gravity which besides the mass $M$, also depends on two parameters $q$ and $\tilde{\alpha}$. We have examined the effects of $q$ and $\tilde{\alpha}$, in a strong-field observation, to the lensing observables due to Charged black holes to $4D$ EGB gravity and compared with those due to Schwarzschild and Reissner-Nordstrom black holes of GR.  We have numerically calculated the strong lensing coefficients $ \bar{a} $ and $ \bar{b}$, and lensing observables $ \theta_{\infty} $, $s$, $r_{\text{mag}}$,  $u_{m}$ as functions of $\tilde{\alpha}$ and $q$ for relativistic images. In turn, we have applied our results to the supermassive black holes, Sgr A* and M87*, at the centre of galaxies.  Interestingly, we find that, $\bar{a} $ increases when we increase $q$ and $\tilde{\alpha}$ whereas $\bar{b}$ and deflection angle $\alpha_D$ decrease, and observe the diverging behavior of deflection angle $\alpha_D$ when $u \to u_{m}$. Besides, for a fixed value of impact parameter, Charged black holes to $4D$ EGB gravity cause a smaller deflection angle as compared to their GR counterparts.

We have also estimated some properties of relativistic images, the variations of $ \theta_{\infty} $, $s$,  and $r_{\text{mag}}$,  as functions of  $\tilde{\alpha}$ and $q$ are depicted in the Figs. \ref{plot7} and \ref{plot8}. We have shown that the angular position of outermost relativistic images $\theta_{\infty} $ and relative magnification of images $r_{\text{mag}}$ are decreasing function of  both $q$ and $\tilde{\alpha}$, but they decrease more sharply with $q$ (cf. Figs. \ref{plot7} and \ref{plot8}), while angular separation between images $s$ increases with both $q$ and $\tilde{\alpha}$. To conclude,  we find that Charged black holes to $4D$ EGB gravity  cause higher angular separation between images but lower magnification than those for the Schwarzschild and Reissner-Nordstrom black holes.

The results presented here are the generalization of previous discussions, on the black holes in GR viz. Schwarzschild  and Reissner-Nordstrom black holes and black holes to $4D$ EGB gravity, and they are encompassed, respectively  in the limits, $\tilde{\alpha}, q \to 0$, $\tilde{\alpha} \to 0$, and $q \to 0$.

\section{Acknowledgments} S.G.G. would like to thank  SERB-DST for the ASEAN project IMRC/AISTDF/CRD/2018/000042 and also IUCAA, Pune for the hospitality while this work was being done. R.K. would like to thank UGC for providing SRF. S.U.I would like to thank SERB-DST for the ASEAN project IMRC/AISTDF/CRD/2018/000042.

\end{document}